# Size effects in multiferroic BiFeO$_3$ nanodots: A first-principles-based study

Wei Ren* and L. Bellaiche

*Physics Department, University of Arkansas, Fayetteville, AR 72701, USA*

An effective Hamiltonian scheme is developed to investigate structural and magnetic properties of BiFeO$_3$ nanodots under short-circuit-like electrical boundary conditions. Various striking effects are discovered. Examples include (a) scaling laws involving the inverse of the dots' size for the magnetic and electric transition temperatures; (b) the washing out of some structural phases present in the bulk via size effects; (c) the possibility of tailoring the difference between the Neel and Curie temperatures, by playing with the size and electrical boundary conditions; and (d) an universal critical thickness of the order of 1.6 nm below which the dots do not possess any long-range ordering for the electrical and magnetic dipoles, as well as, for the oxygen octahedral tiltings.

* E-mail: weiren@uark.edu

Multiferroics can simultaneously exhibit ferroelectricity and magnetic ordering [1]. Such class of materials exhibits a magnetoelectric coupling that is of high technological relevance, since it implies that electrical properties are affected by a magnetic field or, conversely, that magnetic properties can be varied by an electric field. Multiferroics, in their *bulk and film* forms, have been intensively studied since the sixties (see, e.g., Refs. [2-16] and references therein). Consequently, an extensive knowledge of multiferroic bulks and films has been gained and numerous breakthroughs have occurred.

On the other hand, it is only very recently that efforts have been directed towards the physics of *zero-dimensional nanoscale multiferroics*, with, e.g., the successful growth of nanometric $BiFeO_3$ (BFO) particles reported in Refs. [17-20]. These pioneering experimental works mostly focused on magnetic properties of such nanostructures. In particular, they pointed out the existence of magnetism even for the smallest grown nanoparticles (of the order of 4nm-10nm)—in addition to report that the Neel temperature decreases as the size decreases. One may thus wonder if there is any critical size below which magnetism would disappear. Moreover, to the best of our knowledge, the size dependency of the critical (Curie) temperature below which ferroelectricity exists is unknown in multiferroic nanodots, despite the fact that many important responses (such as piezoelectric and dielectric coefficients) are maximal at this temperature. Similarly, the dependency of a physical quantity that plays an important role on electric and magnetic properties of BFO bulks and films [16, 21, 22], that is the oxygen octahedral tilting, on the dot's size is an unexplored area.

The aim of this Report is to reveal the effects of the size on finite-temperature structural and magnetic properties of BFO nanodots being under short-circuit-like electrical boundary conditions, via the use of a first-principles–based scheme. As we will see, several phenomena of fundamental and technological importance are discovered.

Here, we want to accurately mimic *finite-temperature* properties of stress-free BFO cubic nanodots, as modeled by non-periodic $n{\times}n{\times}n$ supercells (where $n$ is an even integer ranging between 2 and 16, and whose product by 4 provides the lateral size of the dot in Angstroem) [23]. For that we use an effective Hamiltonian technique within Monte-Carlo (MC) simulations, for which the degrees of freedom are the local soft modes $u_i$ in unit cells $i$ (whose product with an effective charge, $Z^*$, provides the electrical dipoles in that cell $i$); the *antiferrodistortive* (AFD) $\omega_i$ vectors that characterize the oxygen octahedral tiltings, and whose direction is the axis about which the $FeO_6$ octahedron of cell $i$ tilts while its magnitude provides the angle (in radians) of such tilting; the magnetic dipoles $m_i$ on the Fe sites $i$; and the total strain tensor $\eta=\eta_H+\eta_I$ that incorporates both the homogeneous ($\eta_H$) and inhomogeneous parts ($\eta_I$). The total energy of such effective Hamiltonian is written as:

$$E(\{u_i\},\{\omega_i\},\{m_i\},\{\eta_H\},\{\eta_I\},\beta) = E_{FE-AFD}(\{u_i\},\{\omega_i\},\{\eta_H\},\{\eta_I\}) + \frac{1}{2}\beta\sum_i Z^* u_i \cdot \langle E_{dep} \rangle + E_{MAG-ANI}(\{u_i\},\{\omega_i\},\{m_i\},\{\eta_H\},\{\eta_I\}). \quad (1)$$

where $E_{FE-AFD}$ gathers the energetic terms associated with the ferroelectric, AFD and strain degrees of freedom, as well as their mutual interactions. Its expression is the one given in Ref. [24] for bulk systems, except for the long-range electric dipole-dipole interactions for which we use here the corresponding interactions for zero-dimensional systems under open-circuit (OC) electrical boundary conditions [25]. The second term of Eq. (1) represents the depolarizing energy. It involves the *maximum* depolarizing field $\langle E_{dep} \rangle$ (i.e., the one corresponding to a non-vanishing polarization within ideal OC conditions, and that is self-consistently calculated as in Ref. [25]) and a screening parameter *β* that controls the magnitude of the residual depolarizing field. For instance, *β = 0* corresponds to ideal OC electrical boundary conditions while *β = 1* corresponds to ideal short-circuit (SC) electrical boundary conditions (no residual depolarizing field). Here, we will focus on dots under *SC-like* conditions, that is with *β* slightly smaller or equal to 1. The third term of Eq. (1) involves the magnetic interactions (including their anisotropies) and their couplings with electric dipoles, tilting of oxygen octahedra (including spin canting) and strain. Its analytical expression is the one provided in Ref. [26] for BFO bulk, except for the long-range *magnetic* dipolar interactions for which the corresponding interactions for zero-dimensional systems are used [25]. All the parameters of the total energy of Eq. (1) are extracted from first-principle calculations on relatively small supercells [27, 28]. This total energy is then used in Monte-Carlo simulations with up to 3,000,000 sweeps to equilibrate the system at finite temperature and get converged statistical results. Typical outputs of the simulations for each MC sweep are the supercell average of the local soft modes, *u*; of the ferromagnetic (FM) vector, **M**; of the G-type antiferromagnetic (AFM) vector [22, 27] $L = \frac{1}{N}\sum_i m_i (-1)^{n_x(i)+n_y(i)+n_z(i)}$ ; and of the antiferrodistortive vector associated with the R-point of the cubic Brillouin zone [22, 27] $\omega = \frac{1}{N}\sum_i \omega_i (-1)^{n_x(i)+n_y(i)+n_z(i)}$. Note that these sums run over all the $N$ unit cells of the nanodot and that $n_x(i)$, $n_y(i)$ and $n_z(i)$ are the integers locating the 5-atom unit cell $i$ centered around $r_i = [n_x(i)\mathbf{x} + n_y(i)\mathbf{y} + n_z(i)\mathbf{z}]a$ ($a$ being the predicted 5-atom cubic lattice parameter at 0K, and $\mathbf{x}$, $\mathbf{y}$, $\mathbf{z}$ being the unit vectors along the x-, y- and z-axes, respectively). In the following, we use the `$\langle \cdots \rangle$' notation to indicate statistical averages of these quantities over the MC sweeps. Note that first-principles-based effective Hamiltonian schemes accurately reproduced striking features of BFO bulks and films [29-31], as well as complex structures of low-dimensional ferroelectrics [29-31]. Moreover, effective Hamiltonians have also been previously shown to provide accurate *size dependencies* of properties in nanostructures -- including the disappearance of stripe domains below a three-unit-cell thickness, and the square-root dependency of domain width with thickness (for thicknesses above 3 unit cells), in ferroelectric ultrathin films [29, 32-36].

Let us first investigate a 16×16×16 stress-free BFO nanodot (its lateral size is thus around 6.4 nm). Figures 1 report the temperature evolution of the electric, AFD and magnetic order parameters for ideal SC conditions. Moreover, Figures 2 display the electric, AFD and magnetic configurations of the resulting ground states. This stress-free BFO dot, when under ideal SC conditions, exhibits properties that bear resemblance with those of the corresponding bulk, in the sense that (i) there is a critical Curie temperature (around 1130K [27, 28]), $T_C$, below which electric dipoles homogeneously lie along the [111] pseudo-cubic direction and below which any two neighboring oxygen octahedra tilt in antiphase about [111]; (ii) the dot also possesses another critical temperature (around 685K [27, 28]), $T_N$, below which the 0D-system acquires a G-type antiferromagnetism combined with a weak spin-canting-induced ferromagnetic vector -- with the AFM vector lying near a (111) plane and with the FM vector being nearly perpendicular to both the AFD and AFM vectors; and (iii) above $T_C$, the phase is purely antiferrodistortive (as consistent with the experimental findings of Refs. [37, 38]) until a third critical temperature, $T_{AFD}$, at which the system becomes paraelectric cubic. However, some significant differences also exist between the properties of this dot under SC conditions and those of the bulk. For instance, the purely AFD phase existing above $T_C$ occurs in a much narrower temperature range in the dot than in the bulk: the difference between $T_{AFD}$ and $T_C$ is of the order of 200K in the 6.4nm dot versus 350K in the bulk [27, 28]. In other words, size effects tend to suppress this high-temperature phase that solely exhibits tilting of oxygen octahedra. In fact, this purely AFD phase is numerically found to completely vanish in $n \times n \times n$ nanoparticles with *n* smaller than 10, leading to a direct transition from the paraelectric cubic state to the (rhombohedral) ground-state phase for dots having a lateral size smaller than 4nm.

Let us now investigate *the size dependency* of the $T_C$ and $T_N$ critical temperatures of the $n \times n \times n$ BFO dots for ideal SC conditions, as well as, for $\beta=0.98$ – that represents a point in the short-circuit-like regime having a non-vanishing depolarizing field. Figure 3 provides such information. It indicates that, for *n* larger than 4, these transition temperatures *decrease* as the dot's size decreases, as consistent with the corresponding decrease of the Neel temperature and of the electrical polarization observed in Ref. [19] for BFO nanoparticles. The inset of Fig. 3 confirms another experimental finding of Ref. [19], namely that the magnetic transition becomes more diffuse as the dot shrinks in size.

These predictions further demonstrate the accuracy of our numerical tool. Moreover, Fig. 3 reveals that the predicted critical temperatures follow rather well an *A-B/n* relation – where *A* and *B* are both positive constants for a given transition. Such relation is consistent with experimental and theoretical findings that the Curie temperature of ferroelectric nanowires obeys a $1/d$ scaling law, where *d* is the wire's diameter [39, 40]. Moreover, for the dot under ideal SC electrical boundary conditions ($\beta=1$), $T_N$ decreases slightly with *n* (in agreement with the observations of Ref. [19]) while $T_C$ has a more pronounced variation with the dot's size (the *B* parameter of the scaling law of the ferroelectric transition is around 70% larger than that of the magnetic transition). The coupling between magnetism and polarization is thus rather weak in BFO dots under SC conditions – as similar to the case of the corresponding bulk [41]. A magnetoelectric coupling being weak can also be guessed by comparing the data for $\beta=1$ and $\beta=0.98$. As a matter of fact, the Curie temperature is considerably smaller for $\beta=0.98$ (reflecting the fact that the depolarizing field in the short-circuit-like regime desires to annihilate the polarization), while the Neel temperature is merely unaffected by such change in electrical boundary conditions. Interestingly, the much more pronounced sensitivity of $T_C$ than $T_N$ with size and electrical boundary conditions can be used as a route to bring these two critical temperatures closer to each other. For instance, for *n*=4 and $\beta=0.98$, the difference between $T_C$ and $T_N$ is around 320K, that is around 24% smaller than the *bulk* value of 420K ± 20K. Tuning the size and electrical boundary conditions of the nanodot in the short-circuit-like regimes can therefore lead to a tailoring of some BFO's properties – such as an enhancement of piezoelectric and dielectric responses, as well as the magnetoelectric (ME) coefficients, just below the Neel temperature. For instance, the magnitude of the quadratic ME coefficients is increased by around 30% (respectively, 20%) at 500K (respectively, 300K) in the *n*=6 dot (with β=0.98) with respect to the corresponding values in the bulk.

Moreover, we numerically found that the dot with n=2 has *no* critical temperature for β=1 and β=0.98. More precisely, for the dot with n=2, the local electric dipoles and tiltings of oxygen octahedra fully vanish at any site for any temperature (for any MC sweep). On the other hand, the AFM vector has a magnitude that is still significant at low temperature for any MC sweep. However, it changes of direction between different MC sweeps, resulting in null Cartesian components when averaging over all the MC sweeps. The critical size, below which electric dipoles, AFD displacements and magnetic dipoles can not organize themselves anymore in an ordered fashion, is thus as small as 1.6 nm. Such prediction is consistent with the fact that various experiments [17-20] all reported the existence of magnetism for nanoparticles as tiny as 4nm-10nm. This discovery makes BFO nanodots really attractive for nanotechnology applications.

As shown in Fig. 1(d), our calculations also predicted a weak spin-canting-induced ferromagnetic (FM) vector in the investigated BFO nanodots. Its magnitude is rather small, that is of the order of 0.02-0.03 Bohr magneton per Fe site, for any studied size. Such values are in excellent agreement with some experiments on thin films [42] and in reasonable agreement with some recent measurements on BFO nanoparticles [18]. On the other hand, they contrast with the magnetization of 0.4 Bohr magneton reported in some low-dimensional BFO systems [17]. Such latter large value may be due to a large fraction of uncompensated spins from the surfaces or oxygen deficiencies in the grown sample

[18, 19, 43]. Note that we also conducted computations on $n \times n \times n$ nanodots, with $n$ being an odd integer, since these latter nanoparticles do possess uncompensated magnetic dipoles. We indeed found there that such uncompensated dipoles induce a weak FM moment that increases in magnitude as the dot's size decreases.

We hope that our findings will pave the way to more experimental studies on zero-dimensional multiferroics and prompt further applications in magnetoelectronics and spintronics.

We thank I. Ponomareva for useful discussions. This work is supported by ONR Grants N00014-04-1-0413, N00014-08-1-0915 and N00014-07-1-0825, NSF grants DMR 0701558 and DMR-0080054, and DOE grant DE-SC0002220. Some computations were made possible thanks to the MRI grant 0722625 and to the HPCMO of the DoD.

Captions:

FIG. 1. Temperature dependency of the order parameters in a BFO nanodot of 6.4 nm lateral size, and under ideal short-circuit electrical boundary conditions. Panel (a) shows the Cartesian components of the local mode (that is proportional to the electric polarization). Panel (b) displays the Cartesian components of the antiferrodistortive vector, $\omega$. Panels (c) and (d) represent the magnitude of the antiferromagnetic and ferromagnetic vectors, respectively. The x-, y- and z-axes are chosen along the pseudo-cubic [100], [010] and [001] directions, respectively.

FIG. 2. Three-dimensional patterns at 10K for a BFO nanodot having a lateral size of 6.4 nm, and under ideal short-circuit electrical boundary conditions. Panel (a) shows the electric dipoles. Panel (b) displays the $\omega_i$ antiferrodistortive motions. Panel (c) displays the $m_i$ magnetic dipoles. The x-, y- and z-axes are chosen along the pseudo-cubic [100], [010] and [001] directions, respectively. Only a *3×3×3* inner part of the *16×16×16* dot is shown here, for clarity.

FIG. 3. Size dependency of critical temperatures in *n×n×n* BFO nanodots (whose lateral sizes are nearly *4n*, when given in Angstroems). The filled symbols correspond to ideal short-circuit conditions, while the half-filled symbols show data for nanodots with $\beta=0.98$. Lines represent fittings by *A-B/n* scaling laws. The inset displays the temperature evolution of the magnitude of the AFM vector when $\beta=1$ for the 6×6×6 and 16×16×16 dots. Such inset shows that the magnetic transition becomes more diffuse as the size of the dot decreases (such conclusion is also found for BFO dots with $\beta=0.98$). The uncertainty of the predicted critical temperatures is estimated to be 10K. For each dot, the Neel temperature is identified as the inflection point of the <|**L**|>-versus-temperature curve, while $T_C$ is identified by the jumps of the local modes (see Fig. 1(a)).

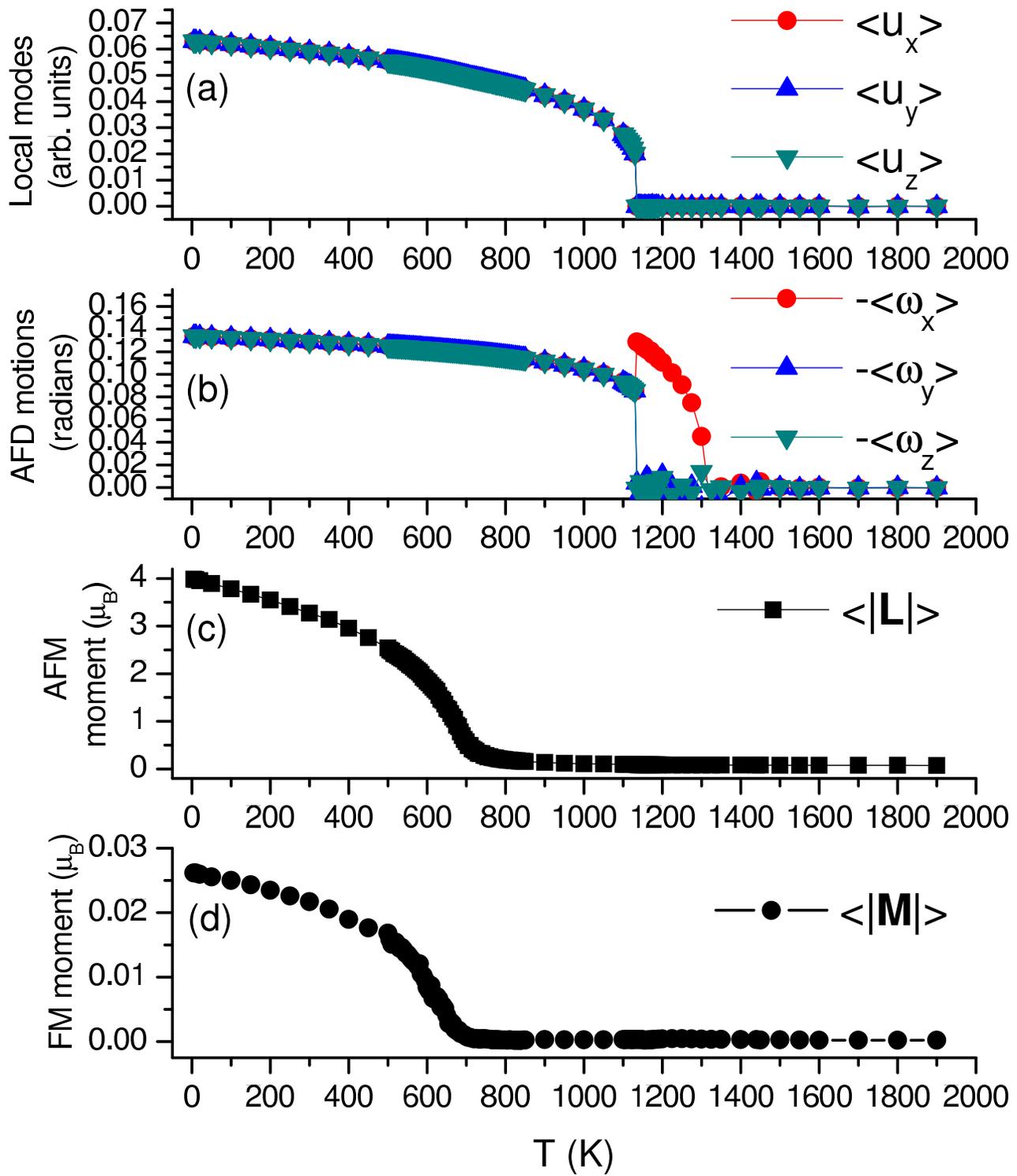

Figure 1    LT12707BJ    09AUG2010

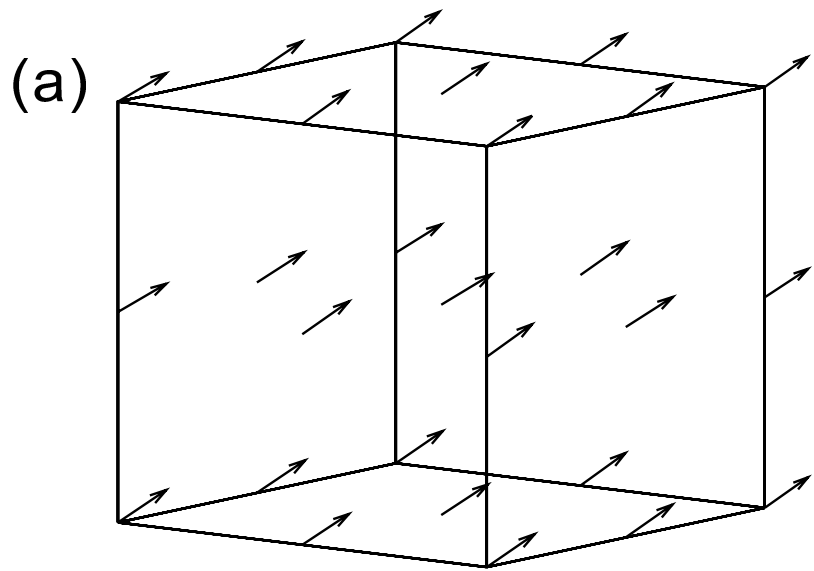

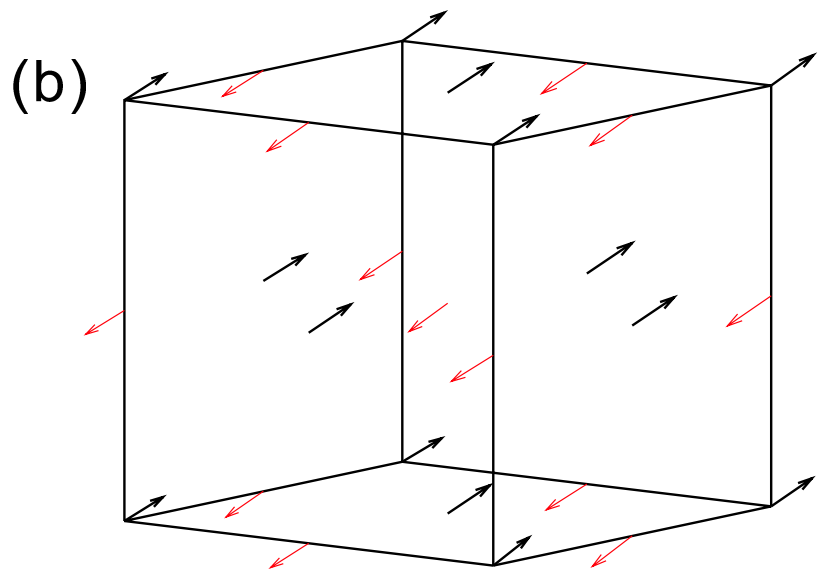

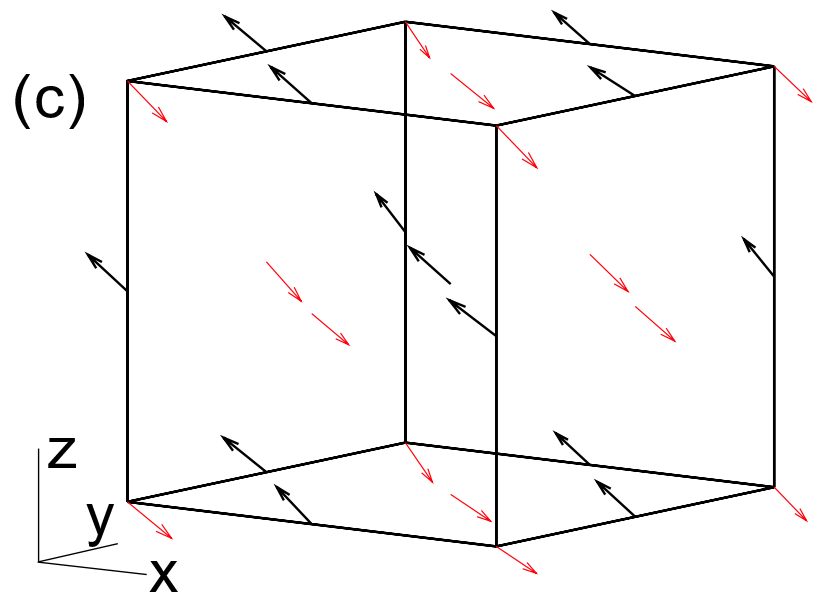

Figure 2    LT12707BJ    09AUG2010

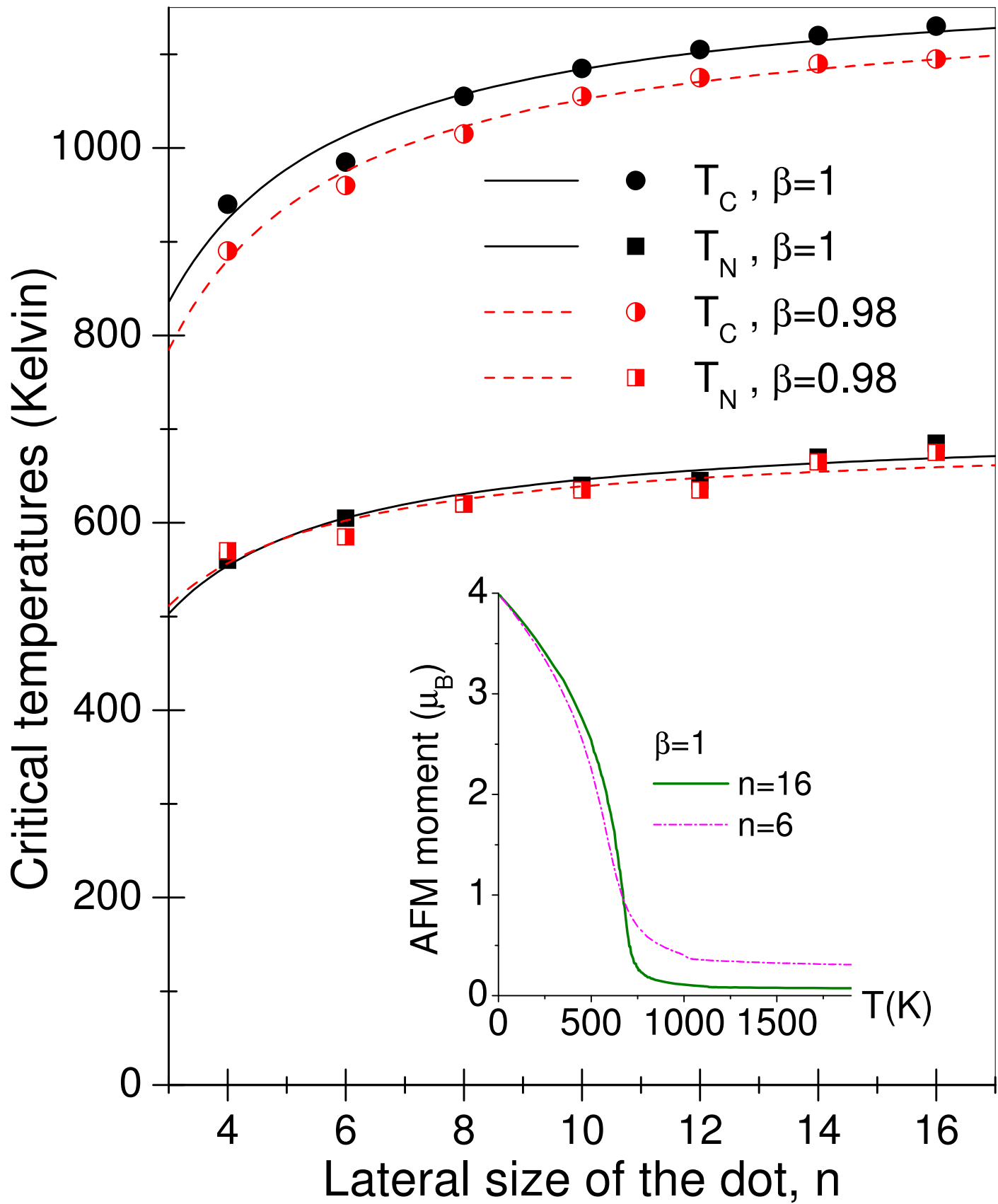

Figure 3     LT12707BJ     09AUG2010